%% file: NPARSEC1.9.tex
\documentclass[useAMS,usenatbib]{mn2e}
\usepackage{graphicx}
\title[NPARSEC I]
{NPARSEC: NTT Parallaxes of Southern Extremely Cool objects. 
Goals, targets, procedures and first results.}

\author[R. L. Smart et al.]
{R. L. Smart$^{1}$\thanks{E-mail: smart@oato.inaf.it},
C. G.     Tinney$^2$,
B.        Bucciarelli$^1$,
F.        Marocco$^3$,
U.        Abbas$^1$,
A.        Andrei$^{4,1}$,\newauthor
G.        Bernardi$^1$,
B.        Burningham$^{3,4}$,      
C.        Cardoso$^1$,
E.        Costa$^5$,
M. T.     Crosta$^1$,
M.        Dapr\'a$^1$,\newauthor
A.        Day-Jones$^{1,5}$,      
B.        Goldman$^6$,
H. R. A.  Jones$^3$,
M. G.     Lattanzi$^1$,
S. K.     Leggett$^7$,\newauthor
P.        Lucas$^3$,      
R.        Mendez$^5$,
J. L.     Penna$^4$,
D.        Pinfield$^3$,
L.        Smith$^3$,      
A.        Sozzetti$^1$,\newauthor
A.        Vecchiato$^1$\\
$^{1}$Istituto Nazionale di Astrofisica, Osservatorio Astrofisico di Torino, Strada Osservatorio 20, 10025 Pino Torinese, Italy\\
$^{2}$School of Physics, University of New South Wales, 2052. Australia\\ 
$^{3}$Center for Astrophysics Research, University of Hertfordshire, Hatfield AL10 9AB,UK\\
$^{4}$Observat\'{o}rio Nacional/MCT,R, Gal. Jos\'{e} Cristino 77,CEP20921-400,RJ, Brazil\\
$^{5}$Universidad de Chile,Camino el Observatorio \# 1515, Santiago, Chile,
Casilla 36-D\\
$^{6}$Max Planck Institute for Astronomy, Koenigstuhl 17,D--69117 Heidelberg,
Germany\\
$^{7}$Gemini Observatory, 670 N. A'ohoku Place, Hilo, HI 96720, USA \\
}
\begin{document}

\include{aas_macros}

\date{}

\pagerange{\pageref{firstpage}--\pageref{lastpage}} \pubyear{2012}

\maketitle

\label{firstpage}

\begin{abstract}

  The discovery and subsequent detailed study of T dwarfs has provided many
  surprises and pushed the physics and modeling of cool atmospheres in
  unpredicted directions. Distance is a critical parameter for studies of these
  objects to determine intrinsic luminosities, test binarity and measure
  their motion in the Galaxy.  We describe a new observational program to
  determine distances across the full range of T dwarf sub-types using the
  NTT/SOFI telescope/instrument combination. We present preliminary results
  for ten objects, five of which represent new distances.

\end{abstract}


\section{Introduction}

The prototype T dwarf was discovered in 1995 as a companion to the nearby M
dwarf star Gl229 \citep{1995Natur.378..463N}. This was rapidly followed by
many discoveries in the near-infrared Two Micron All Sky Survey
\citep[hereafter 2MASS;][]{2006AJ....131.1163S} and the optical Sloan Digital
Sky Survey \citep[SDSS;][]{2000AJ....120.1579Y}.  Once discovered, significant
efforts were undertaken to determine their distances \citep[][hereafter
  TIN03]{2004AJ....127.2948V, 2002AJ....124.1170D, 2003AJ....126..975T} to map
out the lower end of the Hertzsprung-Russell diagram and to constrain
models. These early T-dwarf parallax programs were operating in ``discovery''
mode prioritising new and exciting discoveries at cooler and cooler
temperatures. For this reason, in 2010, the number of faint cool T6-T8 dwarfs
with measured parallaxes was more than double the number of brighter hotter
T0-T5 objects per subclass ($\sim$5 vs $\sim$2, see Figure 2).

The deeper United Kingdom Infrared Deep Sky Survey
\citep[UKIDSS;][]{2007MNRAS.379.1599L} and the Canada-France Brown Dwarf
Survey \citep[CFBDS;][]{2008AA...484..469D} programs increased the number of
known T dwarfs and extended the spectral range to T9. Recently, the
Wide-field Infrared Survey Explorer \citep[WISE;][]{2010AJ....140.1868W} has
extended the range into Y dwarfs and found a significant number of new T
dwarfs.  In 2010, this larger number of confirmed T dwarfs motivated us 
to commence an observational program targeting parallax measurements
spanning the full T dwarf range: the {\it NTT Parallaxes of Southern 
Extremely Cool} objects survey (hereafter NPARSEC).

NPARSEC's original goal was to obtain parallaxes for 65 new brown dwarfs 
to ultimately increase the number of objects with precisely measured
parallaxes to 10 per T dwarf sub-class. 
This would allow us to reduce the uncertainty in spectroscopic 
parallaxes from the current level of 0.4 magnitudes per subclass
\citep{2010AA...524A..38M} down to the level of the estimated
cosmic scatter of 0.2 magnitudes per subclass \citep{2009MmSAI..80..674S}.
In practice, a higher than expected observing efficiency, combined
with a significantly larger pool of bright T dwarfs discovered 
in the WISE survey, have enabled us to increase our NPARSEC survey
sample to 85 targets.

This sample will be used to calibrate the absolute magnitude-spectral type
relation that is critical for determining distances of larger samples used in
the determination of the Initial Mass Function and other statistical
properties of the brown dwarf population. The distances will be used to
discover and characterise peculiar objects and unresolved binaries where the
spectroscopic parallax relations for normal disk brown dwarfs are not
reliable. Distances are also needed to characterise {\bf benchmark systems -
  e.g. systems that provide constraints on physical characteristics such
  as mass, age or metallicity - which,} given the degeneracy between age and
spectral type for brown dwarfs, are crucial to understand the various stages
of T dwarf evolution. The PARSEC program \citep{2011AJ....141...54A}, being
conducted by the same team, is a complementary program for the hotter L brown
dwarfs required to attain a complete picture of sub-stellar objects.

In Section 2 we describe the observational program, in Section 3 we
present the target list, in Section 4 the reduction procedures used and a
discussion of the precision attained, and in Section 5 we publish results from
the first two years of observations and discuss individual objects. Finally in
Section 6 we discuss the future of the program and the relative merits of
visitor and service mode observing for ground-based astrometry, with a view to
the impact of Gaia \citep{2001A&A...369..339P} and other large sky surveys
currently underway.

\section[]{Observational program}

\subsection{Telescope and detector}

\begin{figure}
\includegraphics[width=80mm]{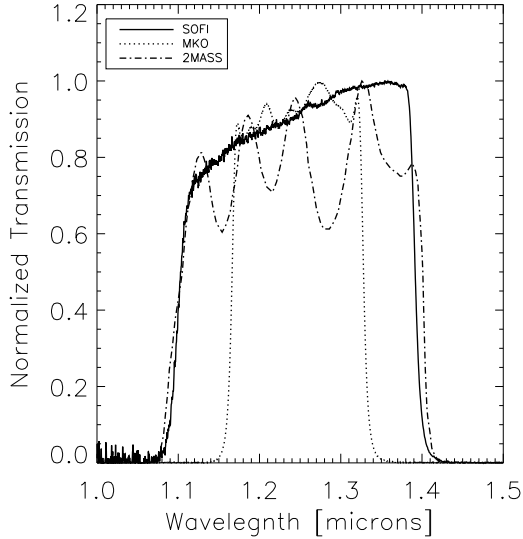}
\caption{The SOFI J passband compared to the MKO and 2MASS bands all
  normalised to the maximum transmission and with no atmospheric absorption. }
\label{filters}
\end{figure}

Observations were carried out using the ESO 3.5m New Technology Telescope
(NTT) and its infrared spectrograph and imaging camera SOFI
\citep{1998Msngr..91....9M}. All observations were carried out in ``Large
Field'' mode, with a pixel scale of 0.288''/pixel and a field of view of 4.9'
x 4.9'. Seeing is rarely better than 0.8”, so the vast majority of images are
well sampled with more than 3 pixels per full-width-at-half-maximum.  All
observations were made in the J band which provides the best compromise
between signal-to-noise and exposure time for these objects. As can be seen
from Figure \ref{filters} the SOFI J band is very similar to the 2MASS J band.
This telescope and instrument combination has a proven track record for
parallax determination \citep[TIN03, ][]{2002AN....323..447N}.

Our goal is to obtain at least 10\% distance precision on all targets.  The
largest distance expected for our targets is $\approx$50\,pc, corresponding to
a parallax of 20\,mas and resulting in a net parallax precision requirement of
2.0\,mas. Many targets -- especially the latest ones -- are much closer with
an average distance of 20\,pc where we will attain a relative precision of
4\%.

The parallaxes and proper motions are determined from the measured coordinates
using the procedures in the Torino Observatory Parallax Program
\citep[hereafter TOPP][]{2003AA...404..317S}.  The parallax is determined from
the equation
\begin{equation}
  \xi_{n,m} = \xi_{n,m_{o}} +
	(t_{m}-t_{m_{o}}) \mu_{n} +
	P_{\xi_{m}}\pi_{n}
\label{xiseq} 
\end{equation}
where: $\xi_{n,m}$ is the position on frame $m$ of star $n$ in gnomic projection standard 
coordinates, $(t_{m}-t_{m_{o}}) $ is the time difference with respect to the
base frame, $P_{\xi_{m}}$ is the parallax factor of observation $m$, and
$\xi_{n,m_{o}}, \mu_{n}$ and $\pi_{n}$ are the base frame position, proper
motion and parallax of the star ${n}$.  If we assume that the observations
have a symmetric distribution of parallax factors we find that the formal error
in the $\xi$ coordinate parallax from the covariance matrix of our observation
equations is given by $\sqrt{\sigma_o^2/(N<{P_{\xi_{m}}^2}>)}$, where N is the
number of observations and $\sigma_o$ the sigma of unit weight of the
least-squares adjustment.  The parallax factor in $\xi$ varies from -1 to 1, hence with
evenly distributed observations the mean $<{P_{\xi}^2}>$ converges to 1/3,
while for the coordinate parallel to declination on average 
converges to 1/5. As
the parallax can be found from both coordinates the formal error simplifies to
$\sigma_{\pi} \approx 1.5\sigma_o / \sqrt{N} $ or by rearranging
$ N \approx 2\sigma_o^2 /\sigma_{\pi}^2$.  To obtain a parallax error of 2 mas with a per-epoch precision floor 
of 6-7mas (see Section ~\ref{sec:precision}) we require 18-24 distinct epochs per target.
We have therefore chosen as a mean goal obtaining 21 epochs per target.


Since the NTT is only operated in visitor mode, we optimised our target list
to be as efficient as possible for full nights. Observations are scheduled on
4 nights spread over a 7 night period, i.e. one 2 night observing block, 3
nights with no observations, then another 2 night observing block.  On each 2
night observing block we attempt to observe all objects on our target list
near the meridian at least once, and then on the following 2 night block we
repeat this sequence.  The ability to split the observations over two nights
allows a larger target list, improves the chances of getting at least one
night without weather problems, and permits us to tailor the target selection
to observing conditions. For example, when seeing is poor we concentrate on
bright targets, or if the wind blows strongly from the north we concentrate on
southern targets. In addition, observations on consecutive
nights are of limited use, as the targets will not have moved significantly
and the general sky conditions and the instrumental setup will probably not
have changed appreciably. However, between the two 2-night blocks our targets
will have significantly moved: assuming an average target with a parallax of
50 \,mas and a proper motion of 500\,mas/yr, the apparent motion will be
13\,mas, i.e. more than twice our nominal precision.

The average night length in La Silla is 10.5 hours so selecting targets evenly
spaced in right ascension we observe on average 10.5/24, or 44\%, of our
targets on each 2-night block.  As we require 21 distinct epochs per target,
we budget 21/2 runs per target, hence a total of 21/2 * 24/10.5 = 24 runs. We
requested one observing run every 6-8 weeks for 3 years for a total of 96
nights which was granted by ESO starting 2010-10-01.  This ``first order''
calculation does not take into account the differing night lengths, or
seasonal weather differences. We have followed this plan over the first two
years, and the number of observations to date (column N$_E$ in Table
\ref{topp01_write}) shows that the summer targets tend to have (on average)
less epochs than the winter targets.  In our current (and last) year of
operation, we will tailor time requests to ``round out'' those objects
requiring remedial treatment.

\subsection{Observation procedures}
During the day we collect darks to cover all possible exposure times. We
obtain dome flats using the SOFI team {\it specialdomeflat} observing block,
and each night we take sky flats. After sky flats are completed we carry out
an image analysis in the area of the first target to configure the NTT's
meniscus mirrors. If the seeing is particularly good, or the images later in
the night particularly elliptical, we redo this image analysis. It is usually
possible to begin target observations after the first image analysis has been
completed even though this is often before nautical twilight. During the
twilight time we concentrate on brighter targets with shorter exposure times
to minimise the effect of the brighter background sky.

The observation procedure starts with a short acquisition exposure and a 
move-target-to-pixel shift to (420,420) -- a point slightly off-center in
the SOFI focal plane which avoids trying to do astrometry at the
boundaries between SOFI's four quadrant read-outs. We then begin a nine-point
dither pattern similar to that adopted in TIN03.  At each dither point we take
an exposure with detector integration time, DIT, repeated NDIT times, and
saved as a single co-added file. The telescope is then dithered and a new
observation begun. This pattern is repeated N times (for N=9, 18 or 27) to
obtain a total integration (based on the target's published J magnitude) of
DIT$\times$NDIT$\times$N, as shown in Table \ref{exposuretimes}.  This set of
discrete exposure time combinations allows us to readily obtain the required
dark frames in advance of each night.  The total exposure times are
conservative estimates to encompass a range of sky conditions and produce a
signal-to-noise of at least 50 for the target in the final co-added image.
Finding charts and observing block files were developed over the first few
observing runs, and have remained substantially unaltered for the whole
campaign.

\begin{table}
 \caption{Exposure times as a function of magnitude}
 \label{exposuretimes}
 \begin{tabular}{ccccc}
  Magnitude & DIT & NDIT & N   & Total Time \\
   J        & (s) &      &       & (min) \\
  \hline
  $<$14.0     &  3     & 20   &    9          &  9  \\
  14. - 15. &  10    &  6   &    9          &  9  \\
  15. - 16. &  20    &  6   &    9          & 18  \\
  16. - 17. &  30    &  4   &    9          & 18  \\
  17. - 18. &  30    &  4   &   18          & 36  \\
  $>$18.      &  30    &  4   &   27          & 54  \\
 \end{tabular}
\end{table}

\subsection{Target scheduling}

Splitting the target list over two nights allows us to tailor the observations
to sky conditions. One driver for this split is the NTT active control of the
primary and secondary mirrors. At the beginning of each night we carry out an
image analysis to attain the best mirror configuration for the current
conditions. This is a costly process, taking 15-20 minutes, and the mirror
shape remains partially a function of the altitude (elevation) of the
telescope. We therefore split the targets into northern and southern groups
and carry out the image analysis at the first target of the group for that
night to obtain a configuration for the average elevation.

Nightly conditions often require us to override this grouping of targets, as
when, for example, the NTT must observe downwind (when wind speeds lie between
12-15\,ms$^{-1}$). Since the wind predominantly blows from the north, our
split on north-south lines is systematically impacted by this
restriction. Operationally, on the first night of a 2-night block we attempt
to observe the northern targets, wind permitting; if this is not possible we
concentrate on southern targets so that (wind permitting) we can observe the
northern group on subsequent nights.

When seeing is particularly poor, the ``cost'' of additional exposures to
achieve a signal-to-noise of 50 for faint targets becomes prohibitive. In these
conditions, we therefore concentrate on bright targets with short exposure
times.  This criterion dominates over other concerns, as the number of bright
targets is such that we cannot pick and choose as we wish to obtain a
north-south split.

\section[]{Target list}
\begin{figure}
\includegraphics[width=80mm]{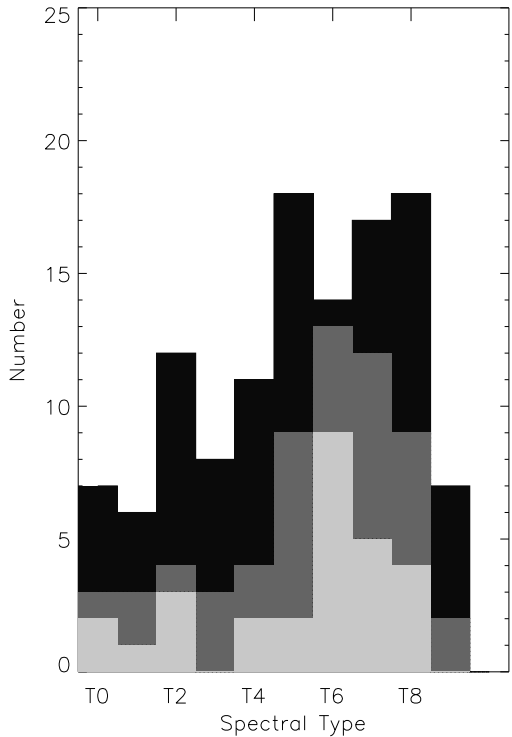}
 \caption{Number of T dwarfs with parallaxes published by 2010 in light grey,
   published today in dark grey and the expected NPARSEC contribution in
   black. }
\label{up4.ecps}
\end{figure}

The target list was chosen from all spectroscopically confirmed T dwarfs
known in 2010 October (supplemented by a few additional late L dwarfs to fill
gaps in the sky coverage). Figure \ref{up4.ecps} shows the spectral type
distribution of T dwarfs with published parallaxes with relative uncertainties
of less than 10\% at three epochs: 2010; today; and that expected at the
conclusion of NPARSEC.

Our original goal was to deliver (in combination with extant published
values), a total of at least 10 objects with measured parallaxes per spectral
sub-class. From Figure \ref{up4.ecps} we can see that this will be possible
for all but the first and last bins. The T0-T1 bins will have additional
objects from the ESO 2.2m PARSEC program \citep{2011AJ....141...54A}, the
Brown Dwarf Kinematics Project \citep[hereafter BDKP,
][]{2012ApJ...752...56F}, the Hawaii Infrared Parallax Program \citep[HIPP,
][]{2012ApJS..201...19D} and the Carnegie Astrometric Planet Search program
\citep[CAPS, ][]{2012ApJ...758...56S}. The last bin and into the Y dwarfs will
be filled by the UKIDSS follow-up parallax program \citep[UFPP,
][]{2010AA...511A..30S} and the various Spitzer \citep{2012sptz.prop90007K,
  2011sptz.prop80233D} and large telescope programs
\citep[e.g. ][]{2012ApJ...759...60T} currently underway.

In Table \ref{topp01_write} we list our targets along with their published J
magnitudes, infrared spectral types, estimated distances and discovery names
from the original 2010 compilation. We used the published J magnitudes in the
2MASS system to derive exposure times. If the published J magnitude was in the
MKO system we use \cite{2004PASP..116....9S} to convert it to the 2MASS
system. This magnitude and the spectral type - absolute magnitude relation of
\citet{2010AA...524A..38M} is used estimate the photometric distances.
\input{topp01_write.tex_edited}

\section[]{Reduction procedures and precision}
\label{sec:precision}
Flats, darks and biases are reduced using standard IRAF routines except for
the dome flats where we use the SOFI team's IRAF script {\it
  special\_flat.cl}.  We co-add the separate dither images into one image per
night using the {\it jitter} routine of the {\it Eclipse}
\citep{1997Msngr..87...19D} version 5.0 package. This routine corrects all
frames using the flat fields, darks and biases.  For each object's dither
sequence it removes bright stars and makes the median of 7 consecutive frames
to use as a sky frame to clean each dither image separately.  Finally it
calculates pixel offsets between each dither and applies them to produce a
re-sampled and co-added final image.

The determination of positions for inclusion in the parallax solution from
this type of sequence can be done in multiple ways. The targets in the HIPP
are sufficiently bright to enable generating one position estimate per
dither. In the UFPP program the centroids are measured from the final combined
image. Finally, the CAPS program uses both approaches, with long integrations
on the main field and a combination of short integrations on a window around
the brighter targets to avoid saturation.

A naive consideration of the errors would suggest that the average of separate
observations will be an improvement of $1/\sqrt{n}$ on the individual
observations. However, this will only be true if each observation can be
treated as an independent sample drawn from a random distribution. The
processes that determine the astrometric precision of a given observation are
complex and correlated within a single night. Therefore there are good reasons
not to expect precision to scale simply with $1/\sqrt{n}$ within a night, nor
even potentially between nights.

When considered as a inter-night problem the number of non-random factors is
large: variable object fluxes, object distributions, atmospheric disturbances,
optical distortions, the detector orientation /  sensitivity will 
all contribute to correlate errors within a night.
These aspects lead to a floor in the astrometric precision which multiple
observations in the same night will not be able to reduce.

To test the precision of the different image treatment procedures we examined
the 27 dither observations of the two targets 0148s72 and 2325s41 on the three
nights 2011-08-19, 2011-11-05 and 2011-11-10.  We also evaluated four
centroiding routines for this data: the two dimensional Gaussian fitting
procedure used in the TOPP; the Cambridge Astrometry Survey Unit’s {\it
  imcore} maximum likelihood barycenter (CASUTOOLS, v 1.0.21); the
SEXTRACTORS’s barycenter; and the SEXTRACTORS’s {\it PSFEX} psf fitting
procedure \citep[v. 2.8.6;]{1996A&AS..117..393B}.

\subsection{Precision: Single Dither Observations}
\begin{figure}
\includegraphics[width=80mm]{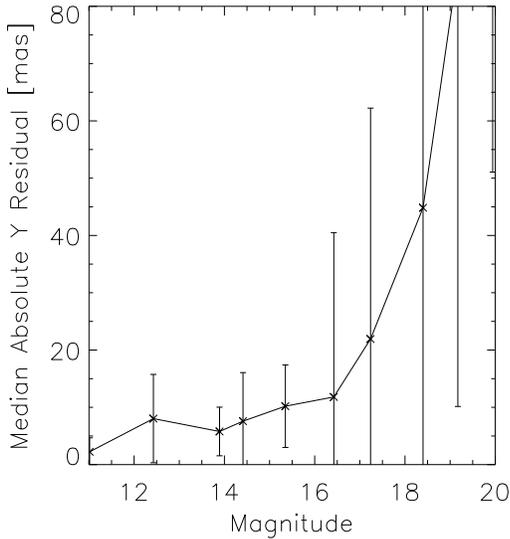}
\caption{Absolute residuals of single dither observations for the field
  0148s72 on the night of 2011-11-10 compared to the coadded image from the
  night of 2011-08-19. The solid line connects the median residuals in equal
  magnitude bins.}
\label{lookatsinglefields_forpaper0148s72W_20111105rr5.ecps}
\end{figure}
First we examine the centroiding precision for single dither images. We fit
the positions from the 27 individual images for each night to the positions in
the co-added image of the targets taken on 2011-08-19. In Figure
\ref{lookatsinglefields_forpaper0148s72W_20111105rr5.ecps} we plot the median
absolute residual in the Y coordinate for 27 frames reduced using the TOPP
centroiding procedures for target 0148s72 on the night of 2011-11-10. The
error bars represent the root-mean-square of the residuals about the median.
At bright fluxes there is a floor to the absolute residual of around
4-8\,mas. At around J=14.5, the residuals slowly rise up to 12\,mas at J=16.5,
and thereafter deteriorate rapidly to 50\,mas at J=18.5.  This deterioration
is due to loss of signal-to-noise for the fainter objects. Similar results are
seen on the other nights and for the other target (2325s41) -- though the
level of the precision floor at bright fluxes did vary significantly from
4\,mas to 10\,mas depending on the observing conditions.  The precisions
obtained in these comparisons was invariant to the centroiding routine used.

\subsection{Precision: Co-added Observations}
\begin{figure}
\includegraphics[width=80mm]{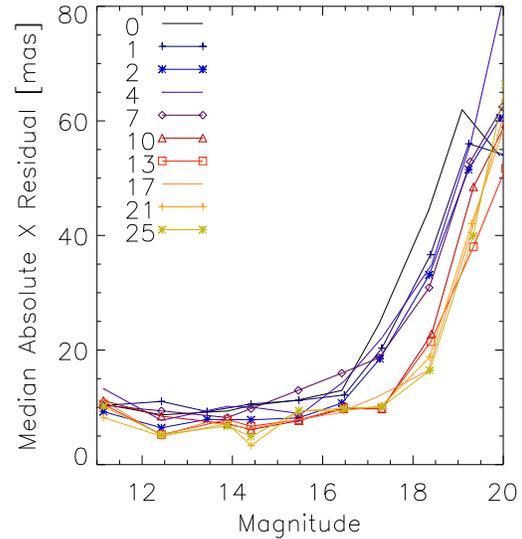}
\caption{Absolute residuals of single dither observations (0 co-adds) and co-added
  combinations as shown in the legends of dithered observations for the field
  0148s72 on the  night of 2011-11-10 compared to the co-added image from
  the night of 2011-08-19. }
\label{lookatcoaddsnallN_forpaper0148s72W_20111105rr5.ecps}
\end{figure}

We now look at how the centroiding precision improves as a function of
co-adding. In Figure \ref{lookatcoaddsnallN_forpaper0148s72W_20111105rr5.ecps}
we compare the single dither images and various co-added combinations for the
0148s72 2011-11-10 observations.  The main improvement obtained is to push
the deterioration in precision to fainter magnitudes, e.g. extending the
precision floor from J=11-15 (obtained with  no co-adding), to J=11-17
(with $>$10 co-added images). Considering only flux for 20 co-added 
images 
we expect an increase of 3 magnitudes, however, this procedure adds
resampling noise so an increase of only 2 magnitudes is reasonable.

\begin{figure}
\begin{center}
\includegraphics[width=80mm]{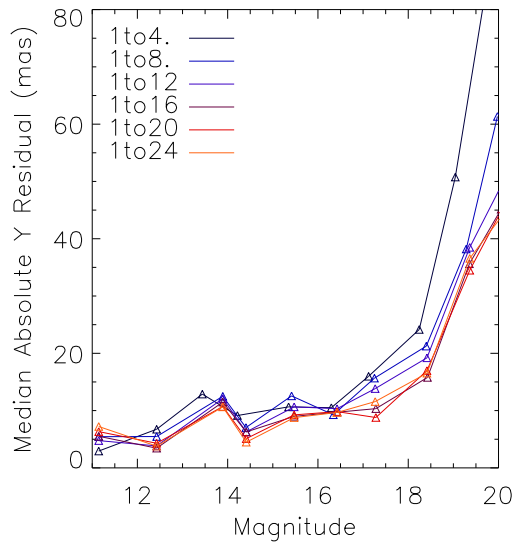}
\caption{As Figure \ref{lookatcoaddsnallN_forpaper0148s72W_20111105rr5.ecps} 
but with first co-added image always frame 1 and the number of co-added 
images shown in the legend. }
\label{lookatcoaddsnfrom1_forpaper0148s72W_20111105rr5.ecps}
\end{center}
\end{figure}

Each co-added combination in Figure
\ref{lookatcoaddsnallN_forpaper0148s72W_20111105rr5.ecps} is made up of
subsets where the first image varied and this may be a source of noise. 
We therefore also followed the change in
precision when the first frame was kept constant. In Figure
\ref{lookatcoaddsnfrom1_forpaper0148s72W_20111105rr5.ecps} the first frame in
the co-add is always frame 1 and the legend indicates the range of frames
co-added. The sense of the results are similar to the previous test but it is
cleaner because the first frame, and related systematic errors, are always in
common. Note, we did not  insist that the comparison objects are common to
all tested sequences and this contributes to the noise seen.

\subsection{Precision: Co-added Observations versus Normal Points}
 
\begin{figure}
\includegraphics[width=80mm]{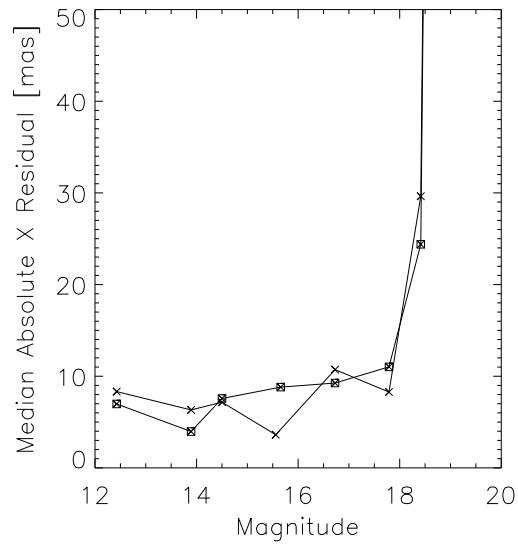}
\caption{A comparison of 27 co-added images (squares) and the averages of the
  separate 27 images (crosses) for the field 0148s72 on the 2011-11-05 both
  compared to the co-added images of the same field on the night 2011-08-19. }
\label{comparecoaddtonormal_forpaperX_26_0148s72W_20111105rr5.ecps}
\end{figure}

We now compare positions from co-added images to averages of the coordinates
from single images. In this test we insist that the comparison objects are
common to both comparisons, i.e.  only the brighter objects detected in the
single images.  In Figure
\ref{comparecoaddtonormal_forpaperX_26_0148s72W_20111105rr5.ecps} for the
field 0148s72 on night 2011-11-05 we compare the positions from 27 co-added
images, and the average of the 27 separate positions to the observation of
2011-08-19. The two methods produce equal precision for the common objects and
this indicates that the error floor is dominated by systematic, rather than
random, errors.

\subsection{Precision: Conclusions}
These tests are only valid for the NTT-SOFI system and reduction procedures
employed here. In particular, we are still experimenting with drizzling and
fractional pixel allocation co-adding procedures.  We conclude that the improvement
from multiple images in the precision floor is very modest, not $1/\surd{n}$,
and the main improvement is to increase the magnitude range for that
floor. This range does not increase in step with signal-to-noise probably
because of noise introduced by the co-adding process. Finally, for the bright
objects, the use of averaged separate observations is not better than using a
co-added image of those separate observations.

In the results published here we use the co-added images to maximize the
magnitude range of the precision floor which we find to be around 6-7\,mas.
The exposures times are selected to ensure that all our targets achieve this
precision. 

\section[]{Preliminary Results}
\begin{figure}
\includegraphics[width=80mm]{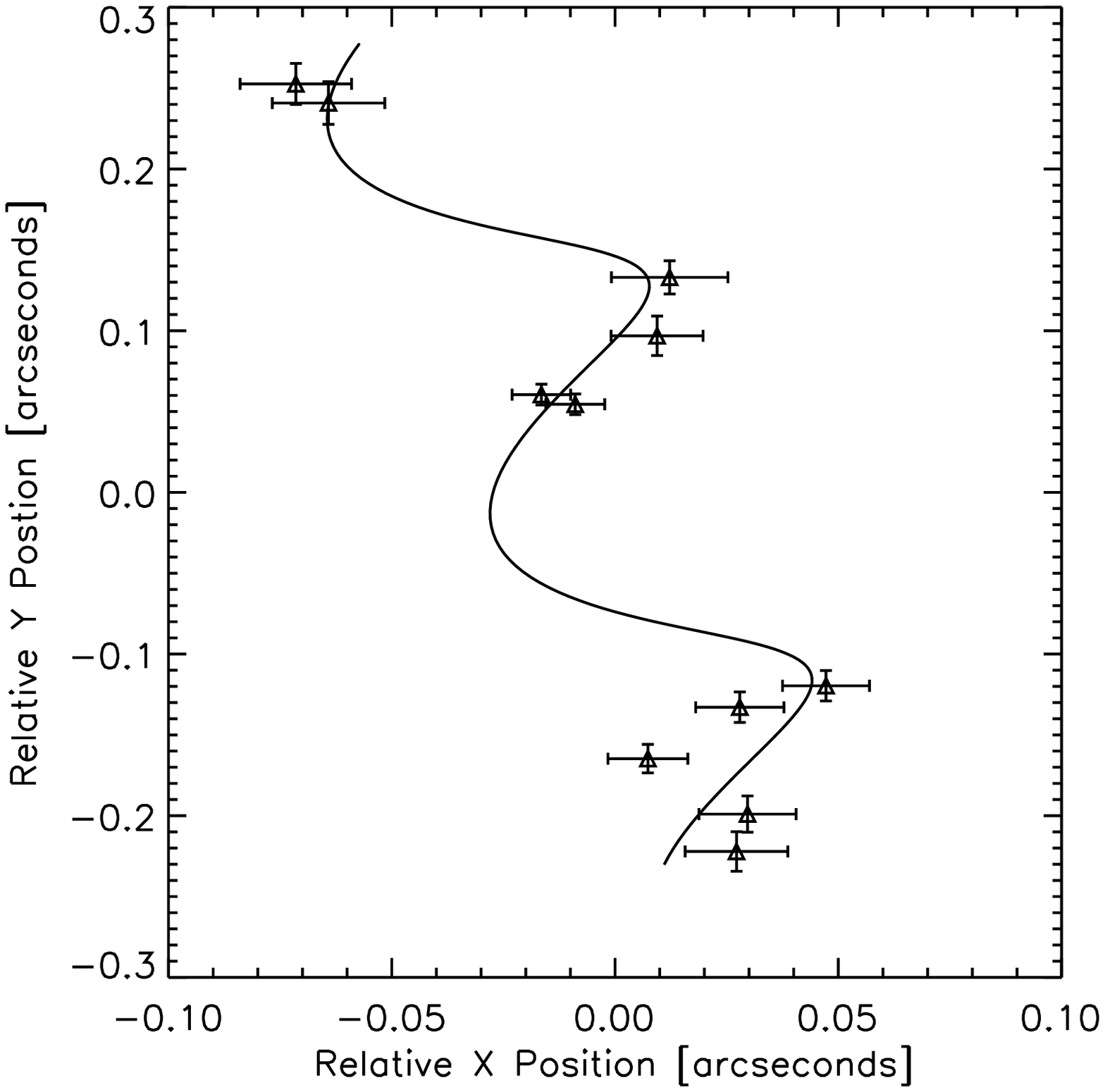}
\includegraphics[width=80mm]{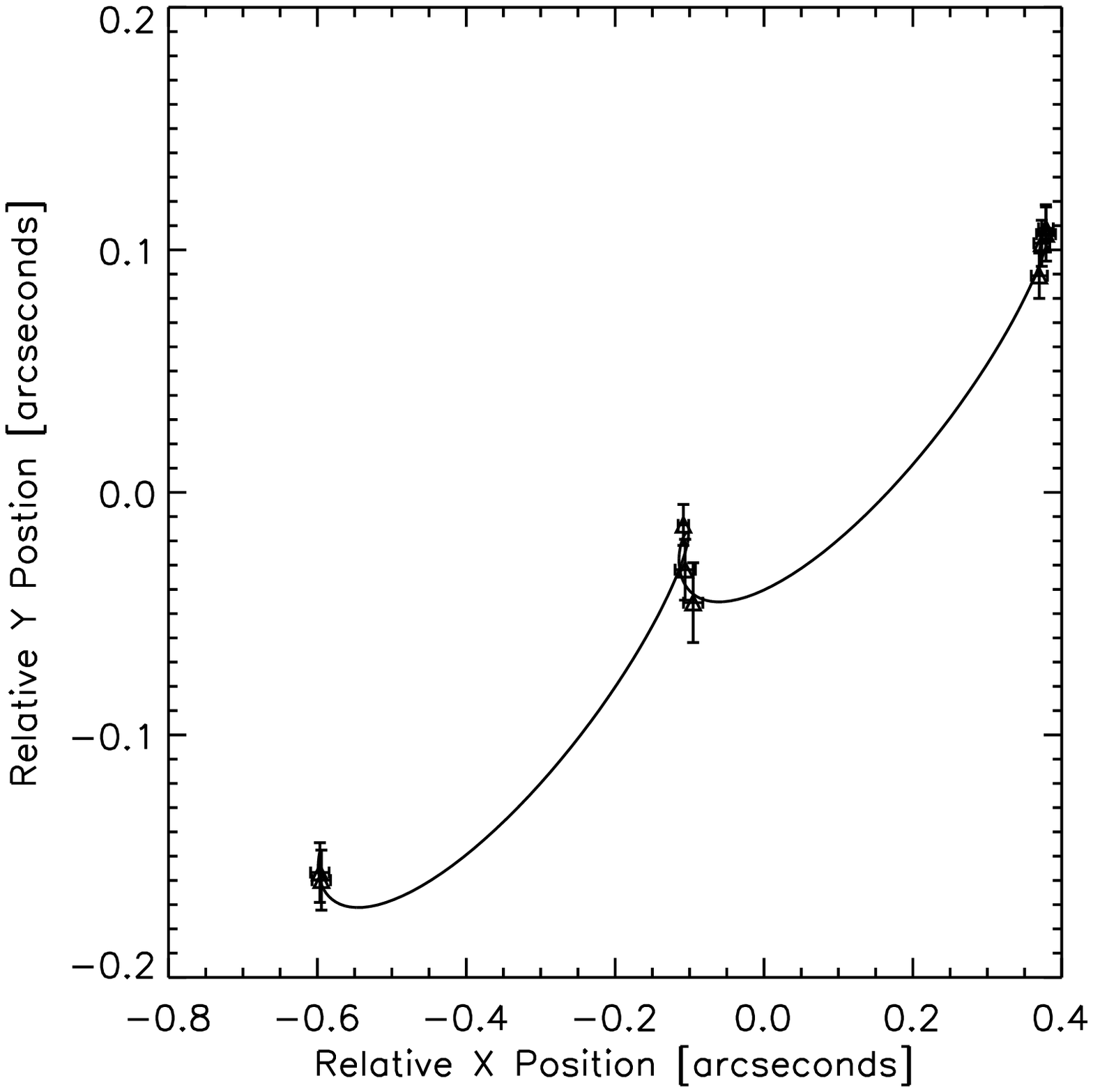}
 \caption{Example target solutions, top panel: 2047s07 with the highest
   error; lower panel: 2139n02 lowest error.}
\label{parallaxes}
\end{figure}

\begin{table*}
\caption{\label{parallax}Preliminary parallaxes and proper motions for NPARSEC targets.}
\centering
\begin{tabular}{lrrrrrrrrrrr}
\hline\hline
NPARSEC & ~~~RA ~~~  & ~~~ Dec ~~~  &  Epoch  &  Absolute $\pi$    & COR  &  $\mu_{\alpha}$~~~~ & $\mu_{\delta}$~~~~&$N_*,N_e$  &$\Delta$T\\
~~Name  & (h:m:s)~~    & ($^\circ$:':'')~~& (yr)~~ &  (mas) ~~      &(mas) &   (mas/yr)   ~   & (mas/yr)   ~     &           & (yr)    \\
\hline  \\                                                                                                                         
0310n16 &  3:11:00.0 & +16:48:15.3 & 2011.64 &    36.9 $\pm$  3.4 & 1.58 &   245.9 $\pm$  4.0 &     6.2 $\pm$  3.3 &  78, 12 &   1.89    \\
1828s48 & 18:28:36.0 & -48:49:03.7 & 2012.58 &    87.9 $\pm$  2.0 & 0.83 &   234.6 $\pm$  2.5 &    88.3 $\pm$  2.6 & 278, 13 &   1.91    \\
2043s15 & 20:43:17.7 & -15:51:04.3 & 2011.62 &    22.8 $\pm$  4.7 & 1.32 &    43.7 $\pm$  4.8 &  -109.4 $\pm$  3.1 & 173, 12 &   1.91    \\
2047s07 & 20:47:49.5 & - 7:18:21.8 & 2012.58 &    33.2 $\pm$  5.5 & 1.47 &    35.9 $\pm$  5.2 &  -241.1 $\pm$  5.0 & 155, 11 &   1.91    \\
2139n02 & 21:39:27.1 & + 2:20:23.9 & 2012.44 &   101.5 $\pm$  2.0 & 1.36 &   485.9 $\pm$  2.0 &   124.8 $\pm$  2.7 &  86,  9 &   1.91    \\
2151s48 & 21:51:38.8 & -48:53:56.5 & 2010.76 &    60.0 $\pm$  3.8 & 1.61 &   414.7 $\pm$  3.5 &  -201.7 $\pm$  4.5 &  52,  9 &   1.90    \\
2154s10 & 21:54:25.1 & -10:23:01.6 & 2010.75 &    37.2 $\pm$  3.5 & 1.62 &   258.0 $\pm$  3.1 &    63.3 $\pm$  5.3 &  54, 11 &   1.91    \\
2228s43 & 22:28:29.0 & -43:10:30.4 & 2011.85 &    92.1 $\pm$  2.6 & 1.46 &    51.7 $\pm$  3.2 &  -301.3 $\pm$  1.6 &  18, 11 &   1.91    \\
2342n08 & 23:42:28.9 & + 8:56:20.0 & 2010.75 &    34.3 $\pm$  5.1 & 1.59 &   264.1 $\pm$  4.4 &   -52.6 $\pm$  3.0 &  56, 13 &   1.91    \\
2356s15 & 23:56:54.3 & -15:53:18.5 & 2010.75 &    57.9 $\pm$  3.5 & 2.21 &  -430.3 $\pm$  5.8 &  -607.9 $\pm$  3.0 &  35,  8 &   1.67    \\
\hline \\
\end{tabular}

COR = correction to absolute parallax, $N_*$ = number of
reference stars, $N_e$ = number of epochs, $\Delta$T = epoch 
range.
\end{table*}

In Table \ref{parallax} we report preliminary results from the first two years
of observations for 10 of our targets. The parallax determination procedures
are substantially the same as those used in \cite{2003AA...404..317S}. There
are two differences in the application for this program: (I) We do not correct
for differential colour refraction as we are working in a wavelength region
where this effect is negligible \citep{2011AJ....141..117J}.  (II) The
correction from relative to absolute parallax (COR in Table \ref{parallax}) is
calculated using the model of Mendez \& van Altena (1996\nocite{men96})
transformed into the J band which we conservatively estimate to have an error
of 30\% \citep{1997ASSL..212..195S}.  Five of the objects in Table
\ref{parallax} overlap with published targets (see Table
\ref{publishedvalues}). Our results are in reasonable agreement, with some
exceptions which are discussed in the following sub-sections.

\begin{table}
 \caption{Objects with previously published parallaxes}
 \label{publishedvalues}
 \begin{tabular}{ccccc}
NPARSEC  & NPARSEC     & Literature  & Reference  \\
~~Name & Parallax      & Parallax    &            \\
  \hline          
1828s48 & 87.9$\pm$2.0 & 84.0$\pm$ 8.0 & BDKP \\
2047s07 & 33.2$\pm$5.5 & 49.9$\pm$ 7.9 & BDKP \\
2151s48 & 60.0$\pm$3.8 & 50.4$\pm$ 6.7 & BDKP \\
2228s43 & 92.1$\pm$2.6 & 94.0$\pm$ 7.0 & BDKP \\
2356s15 & 57.9$\pm$3.5 & 69.0$\pm$ 3.0 & BDKP \\
 ``     &   ``         & 74.4$\pm$ 5.8 & \cite{2004AJ....127.2948V} \\
 \end{tabular}
\end{table}

\subsection{0310n16 (2MASSWJ0310599+164816)} 

This object was seen as a binary in HST and VLT NACO observations
\citep{2010A&A...516A..37S} with a separation of ~200\,mas. In Stumpf at
al. they adopted a spectroscopic distance of 25 pc and assuming a face on
circular orbit this would imply a 5.2 AU separation. From the change in
position angle of 15.5 degrees they found a period of 72 years and hence a
minimum system mass of $\>$30 $M_{Jup}$.  This low mass would imply a very
young system, as an older system would have to be more massive to still be
close to the L/T boundary. However, the hypothesis of a young age is not
supported by a comparison of the space velocities to the locus of known young
objects as done in Marocco et al (2010), and in an examination of the
\cite{2007ApJ...659..655B} spectrum we do not see the triangular H band, nor a
significant H and K band flux enhancement, that we expect from young objects
\cite[e.g. ][]{2001MNRAS.326..695L}. This low mass, and hence young age,
constraint will be relaxed if the orbit is larger than estimated above, which
will be the case if the system is inclined or if the distance is larger than
assumed by Stumpf et al.  Our trigonometric distance measurement is
27.1$^{+3.7}_{-2.3}$\,pc, larger than the adopted 25pc but not enough to change
the overall conclusion.

\subsection{1828s48 (2MASSJ18283572-4849046)}
This object is in the BDKP program but observed over a period before NPARSEC
began. We find that our parallax and proper motions (87.9$\pm$2.0\,mas,
234.6$\pm$2.5\,mas/yr, 88.3$\pm$2.6\,mas/yr) agrees with those found in the BDKP
(83.7$\pm$7.7\,mas, 231.4$\pm$ 0.5\,mas/yr, 52.4$\pm$10.9\,mas/yr) with the
exception of the declination proper motion. This could be due to the short
epoch coverage for both programs (1.91\,yr for NPARSEC and 1.88\,yr for BDKP).
The difference could also be due to orbital motion if the object turns out to
be binary. However, this is unlikely as the photometric distance is the same
as the astrometric one so the object does not appear over luminous.

\subsection{2047s07 (SDSSJ204749.61-071818.3)}
In the BDKP this object was observed on 6 nights over 1.34 years and they
found a parallax and proper motion of: 49.9$\pm$7.9\,mas, 48.7$\pm$11.2\,mas/yr,
-193.8$\pm$11.2\,mas/yr while in NPARSEC with 11 nights over 1.91 years we find
33.2$\pm$5.5\,mas, 35.9$\pm$5.2\,mas/yr, -241.1$\pm$5.0\,mas/yr.  This object has
a 2MASS J magnitude of 16.95 so is at the faint end of the exposure bin for
J=16-17 in Table \ref{exposuretimes} hence the signal-to-noise was often close
to the minimum acceptable. As shown in Figure \ref{parallaxes} the solution is
(not unexpectedly) noisy. Notwithstanding this lower precision the
fit appears well constrained and we note the photometric distance is 29$\pm$5\,pc
which supports the NPARSEC value.

\subsection{2139n02 (2MASSJ21392676+0220226)}
\cite{2009AJ....137....1F} find a proper motion of +507$\pm$22 mas/yr,
123$\pm$22\,mas/yr, consistent with the values found in this program. In
Radigan et al. (2012) \nocite{2012ApJ...750..105R} and Khandrika et al. (2013)
\nocite{2013AJ....145...71K} they find evidence of a 7 hour J band variability
that we will search for when we have a larger observational dataset.  In
\cite{2010ApJ...710.1142B} they indicate this object as a strong binary L/T
candidate; however, based on our distance, it does not appear to be
particularly bright and the residuals to the solution shown in Figure
\ref{parallaxes} show no signatures of orbital motion, hence these
observations do not support the binarity hypothesis. In addition Radigan et
al. (2012) ruled out binarity at distances beyond 1.56 AU level using HST/NICMOS imaging
and Khandrika et al. (2013) do not find any significant variations in radial
velocities using Gemini/NIRSPEC observations.

\subsection{2342n08 (ULASJ2342+0856)} 
This object was provided internally from the UKIDSS T-dwarf search and
spectroscopically typed as a T6.5. In \cite{2010A&A...515A..92S}, based on
photometry, this target was indicated as a probable T7 with a proper motion of
$+229\pm$55 and $-9\pm$9\,mas/yr. These proper motions are inconsistent with
the NPARSEC values of 264.1$\pm$4.4, -52.6$\pm$3.0\,mas/yr but in the
discussion of Scholz he estimates the expected error of this object to be
48mas/yr (see his Table 5), using this value the two estimates become formally
consistent.

\subsection{2356s15 (2MASSIJ2356547-155310)}
The NPARSEC parallax (57.9$\pm$3.5\,mas) is significantly smaller than the
published values (69.0$\pm$ 3.0 \,mas and 74.4$\pm$ 5.8\,mas) while the NPARSEC
proper motions (-430.3$\pm$5.8\,mas/yr, -607.9$\pm$3.0\,mas/yr) are in the
middle of inconsistent published values (-422.7$\pm$4.0\,mas/yr, -615.9$\pm$3.6
mas/yr BDKP and -443$\pm$2\,mas/yr, -600$\pm$2\,mas/yr
\citealt{2004AJ....127.2948V}). This target has the shortest NPARSEC temporal
coverage (1.61 yr) and the lowest number of observations (8) in this
sample so we will wait for more observations before investigating further or
drawing conclusions on the difference between the published and NPARSEC
values.
        
\section[]{General Program Considerations}
\subsection{Visitor vs Service Programs}
It is useful to compare the NPARSEC visitor program to a service program 
such the UFPP. Service programs provide two paths to increase the 
value of  observations for a parallax determination:  a flexibility 
to micro-manage scheduling, and an ability to match observations to
conditions. 
 
Scheduling flexibility in parallax observations usually translates into a
request for observations at twilights when the parallax factor is an absolute
maximum in ecliptic longitude, but, this is also the point where the parallax
factor in the ecliptic latitude is a minimum.  This is justified because the
parallax factor in longitude has a larger range; the factor in latitude being
usually visible at only one extreme and modulated by the sine of the latitude.  However,
modern programs use both coordinates to determine parallaxes
so this is not a particularly strong benefit. Another plus from
scheduling flexibility is, if the program is ranked high enough, to obtain a
more uniform observation distribution. In NPARSEC there are seasons when we
only have 2 observations of a target; in the UFPP, where our program has a
high priority, we obtain over 90\% of the observations requested.

Having conditions that match our requirements is undeniably an advantage.
Also experienced service observers will be more efficient than the frequently 
changing NPARSEC observer, but, this is balanced by greater familiarity of
the NPARSEC observer with the program. For the UFPP we have observations 
that have simple and objective constraints and all the observations to date 
have been useful. 

A benefit in the UFPP programs is the automated reduction pipeline which has
been developed as part of the UKIDSS and has used literally thousands of
images to calibrate the detector. The result is a robust and precise pipeline
that it is impossible to compete with using the inhomogeneous and less
structured SOFI archive.

Both types of programs have different pros and cons and the final efficiency
and precision will probably be dominated by the differences in the
telescope/instrument combinations. The SOFI focal plane is on axis with
0.288'' pixels, has been stably mounted for 15 years on the NTT which has the
most advanced active optics system on a 4m-class telescope. The WFCAM focal
plane is off axis with 0.4'' pixels, has only been mounted for 6 years, but,
it is a fixed instrument on the longest running 4m IR telescope.  Since the
parallaxes of brown dwarfs will remain the domain of ground based programs for
the foreseeable future, this is an interesting comparison.

\subsection{Large Surveys and Future Space missions}
Pan-STARRS \citep{2002SPIE.4836..154K} and the future LSST surveys
\citep{2009arXiv0912.0201L}, will image the sky many times a year,
automatically providing the observations for parallax determination.  However,
surveys are managed to maximise coverage and depth; if this is done at the
expense of astrometric precision or the scheduling flexibility that parallax
observations require, the final precision will suffer. The 1-2\,mas precision
goal we have set for NPARSEC will be challenging for these large surveys,
though even lower precisions with a significantly larger sample is exciting.

The impact of future astrometric space missions is promising but uncertain.
The Japan Astrometry Satellite Mission for Infrared Exploration
\citep[JASMINE, ][]{2002aprm.conf....5G} will measure parallaxes for thousands
of T dwarfs, but the future of that mission is not clear and the precursor,
Nano-JASMINE \citep{2005tvnv.conf..491K}, will not observe any T dwarfs. The
Gaia mission will determine parallaxes with errors better than 0.5\,mas for
{\bf all objects to Gaia magnitude 20 \citep{2012Ap&SS.341...31D}, but the only T
dwarfs this bright are Epsilon Indi Ba and WISE J104915.57-531906.1AB
\citep{2013ApJ...767L...1L}.} Gaia will discover many T dwarfs indirectly as
companions of brighter objects providing an unprecedented wealth of benchmark
systems. Gaia will provide a more precise correction from the
relative-to-absolute parallax for ground based parallax programs; however, as
shown in Table \ref{parallax}, this correction is small, so the overall
precision is dominated by the internal random error. Gaia will
also provide an accurate reference frame which, in theory, could be used to
model focal plane distortions. However, providing the focal plane does not
change over the observational program, the differential nature of ground based
programs uses the full precision of the observations while any modeling will
introduce errors even with a perfect reference catalog, and the availability
of improved Gaia accuracy may not translate into improved parallax estimates.

\section*{Conclusions and Future Work}
We have described the NPARSEC program targets, instrumentation, procedures and
produced first parallaxes and proper motions for ten objects including five
new objects.  The observations are currently scheduled until late 2013 at
which point we will publish a complete set of results.  We are already finding
surprises for individual targets, and we are gathering the supporting
photometric and spectroscopic information to produce a complete homogenous
observational data set to characterise the overall population.

\section*{Acknowledgments}
The authors would like to acknowledge the Marie Curie 7th European Community
Framework Programme grant n.236735 {\it Parallaxes of Southern Extremely Cool
  objects} (PARSEC) International Incoming Fellowship and grant n.247593 {\it
  Interpretation and Parameterisation of Extremely Red COOL dwarfs} (IPERCOOL)
International Research Staff Exchange Scheme.

The main observational data was collected at the European Organisation for
Astronomical Research in the Southern Hemisphere, Chile program 186.C-0756.
Observational data from three other programs have been used to support this
research: A22TAC\_96 on the Italian Telescopio Nazionale Galileo, U09B14 on the
United Kingdom Infrared Telescope and SO2012B-018 on the Southern
Astrophysical Research Telescope. 

We thank the anonymous referee for their useful comments.

\bibliographystyle{mn2e} 
\bibliography{refs}


\label{lastpage}

\end{document}

%% file: aas_macros.tex
%
%
%
%


\def\aj{\rm{AJ}}                   
\def\araa{\rm{ARA\&A}}             
\def\apj{\rm{ApJ}}                 
\def\apjl{\rm{ApJ}}                
\def\apjs{\rm{ApJS}}               
\def\ao{\rm{Appl.~Opt.}}           
\def\apss{\rm{Ap\&SS}}             
\def\aap{\rm{A\&A}}                
\def\aapr{\rm{A\&A~Rev.}}          
\def\aaps{\rm{A\&AS}}              
\def\azh{\rm{AZh}}                 
\def\baas{\rm{BAAS}}               
\def\jrasc{\rm{JRASC}}             
\def\memras{\rm{MmRAS}}            
\def\mnras{\rm{MNRAS}}             
\def\pra{\rm{Phys.~Rev.~A}}        
\def\prb{\rm{Phys.~Rev.~B}}        
\def\prc{\rm{Phys.~Rev.~C}}        
\def\prd{\rm{Phys.~Rev.~D}}        
\def\pre{\rm{Phys.~Rev.~E}}        
\def\prl{\rm{Phys.~Rev.~Lett.}}    
\def\pasp{\rm{PASP}}               
\def\pasj{\rm{PASJ}}               
\def\qjras{\rm{QJRAS}}             
\def\skytel{\rm{S\&T}}             
\def\solphys{\rm{Sol.~Phys.}}      
\def\sovast{\rm{Soviet~Ast.}}      
\def\ssr{\rm{Space~Sci.~Rev.}}     
\def\zap{\rm{ZAp}}                 
\def\nat{\rm{Nature}}              
\def\iaucirc{\rm{IAU~Circ.}}       
\def\aplett{\rm{Astrophys.~Lett.}} 
\def\apspr{\rm{Astrophys.~Space~Phys.~Res.}}
\def\bain{\rm{Bull.~Astron.~Inst.~Netherlands}} 
\def\fcp{\rm{Fund.~Cosmic~Phys.}}  
\def\gca{\rm{Geochim.~Cosmochim.~Acta}}   
\def\grl{\rm{Geophys.~Res.~Lett.}} 
\def\jcp{\rm{J.~Chem.~Phys.}}      
\def\jgr{\rm{J.~Geophys.~Res.}}    
\def\jqsrt{\rm{J.~Quant.~Spec.~Radiat.~Transf.}}
\def\memsai{\rm{Mem.~Soc.~Astron.~Italiana}}
\def\nphysa{\rm{Nucl.~Phys.~A}}   
\def\physrep{\rm{Phys.~Rep.}}   
\def\physscr{\rm{Phys.~Scr}}   
\def\planss{\rm{Planet.~Space~Sci.}}   
\def\procspie{\rm{Proc.~SPIE}}   

\let\astap=\aap
\let\apjlett=\apjl
\let\apjsupp=\apjs
\let\applopt=\ao

%% file: topp01_write.tex_edited
\begin{table*}                                                                                                                                                    
\label{topp01_write}                                                                                                                                              
\begin{minipage}{180mm}                                                                                                                                           
  \caption{NPARSEC Targets. The J is the published J magnitude
    available when the program started in the 2MASS system unless indicated. The
    photometric distance, d$_{p}$, is calculated via the relation between
    absolute magnitude and spectral type provided in
    \citet{2010AA...524A..38M} and the J magnitude.  
}
\begin{tabular}{lllllllllll}                                                                                                                 
Short   & ~~~~~~ Discovery         & Ref &~~~ Discovery $\alpha,\delta$ & ~~J $\pm$ $\sigma$ & Ref & d$_{p} \pm  \sigma$& ~IR & Ref &N$_E$ & $\Delta$T  \\
Name    & ~~~~~~~~ Name              & Dis.&~~~~~  hrs,deg J2000      &                     & ~J     &   ~~pc   & SpT       & SpT &      &   yr    \\ 
0024n00 &     ULASJ002422.94+002247.9 &  18 &   0.4063721,  0.37997222 &  18.16 $\pm$ .07 & 30$^a$ &  55$\pm$8  & T4.5       & 18 &     9 & 1.90 \\
0034n05 &      2MASSJ00345157+0523050 &   8 &   0.5809916,  5.3847222 &   15.53 $\pm$ .05 & 13 	   &  12$\pm$2  & T6.5	    & 14 &    10 & 1.90 \\
0050s33 &      2MASSJ00501994-3322402 &  11 &   0.8388722,  -33.377833 &  15.93 $\pm$ .07 & 13 	   &  12$\pm$2  & T7	    & 14 &    11 & 1.90 \\
0136n09 &     IPMSJ013656.57+093347.3 &  15 &   1.6157291,  9.5631444 &   13.45 $\pm$ .03 & 13 	   &   6$\pm$1  & T2.5	    & 15 &     9 & 1.90 \\
0138s03 &   WISEPJ013836.59-032221.28 &  28 &   1.6434972,  -3.3725778 &  16.39 $\pm$ .10 & 13 	   &  25$\pm$4  & T3p	    & 28 &     8 & 1.66 \\
0148s72 &   WISEPJ014807.25-720258.75 &  28 &   1.8020139,  -72.049653 &  18.96 $\pm$ .07 & 28 	   &   9$\pm$1  & T9.5	    & 28 &     9 & 1.68 \\
0203s01 &     SDSSJ020333.26-010812.5 &   9 &   2.0592168,  -1.1366389 &  17.69 $\pm$ .04 & 30$^a$ &  31$\pm$4  & L9.5       &  9 &    10 & 1.90 \\
0223s29 &   WISEPJ022322.34-293258.23 &  28 &   2.3895388,  -29.549508 &  17.34 $\pm$ .15 & 28	   &  14$\pm$2  & T8	    & 28 &     7 & 1.66 \\
0247s16 &     SDSSJ024749.90-163112.6 &  13 &   2.7971611,  -16.520333 &  17.19 $\pm$ .18 & 13	   &  36$\pm$6  & T2$\pm$1.5 & 13 &    10 & 1.90 \\
0254n02 &    WISEPJ025409.45+022359.1 &  27 &   2.9026251,  2.3997611 &   16.56 $\pm$ .16 & 27	   &  10$\pm$2  & T8	    & 28 &     6 & 0.81 \\
0255s47 &           DENIS-PJ0255-4700 &   1 &   2.9176584,  -47.014136 &  13.25 $\pm$ .03 & 13	   &   4$\pm$1  & L9	    & 14 &     9 & 1.50 \\
0310n16 &       2MASSWJ0310599+164816 &   3 &   3.1832943,  16.804306 &   16.02 $\pm$ .08 & 13	   &  16$\pm$2  & L9	    & 14 &    12 & 1.89 \\
0325n04 &     SDSSJ032553.17+042540.1 &  13 &   3.4314499,  4.4279444 &   16.25 $\pm$ .14 & 13	   &  20$\pm$3  & T5.5	    & 13 &    10 & 1.90 \\
0329n04 &              ULASJ0329+0430 &  29 &   3.4889500,  4.5068056 &   17.50 $\pm$ .03 & 30$^a$ &  39$\pm$6  &  T5        & 29 &     8 & 1.89 \\
0348s60 &      2MASSJ03480772-6022270 &   7 &   3.8021443,  -60.374169 &  15.32 $\pm$ .05 & 13 	   &   9$\pm$1  & T7	    & 14 &     9 & 1.17 \\
0407n15 &      2MASSJ04070885+1514565 &   8 &   4.1191249,  15.249028 &   16.06 $\pm$ .09 & 13 	   &  20$\pm$3  &  T5	    & 14 &    10 & 1.17 \\
0415s09 &       2MASSIJ0415195-093506 &   5 &   4.2554278,  -9.5851667 &  15.69 $\pm$ .06 & 13 	   &   6$\pm$1  & T8	    & 14 &    11 & 1.33 \\
0423s04 &    SDSSpJ042348.57-041403.5 &   6 &   4.3968277,  -4.2343056 &  14.47 $\pm$ .03 & 13 	   &   9$\pm$1  & T0	    & 14 &     9 & 1.58 \\
0510s42 &      2MASSJ05103520-4208140 &  17 &   5.1764555,  -42.137406 &  16.22 $\pm$ .09 & 13 	   &  22$\pm$3  & T5	    & 17 &    14 & 1.35 \\
0516s04 &      2MASSJ05160945-0445499 &   7 &   5.2692919,  -4.7638611 &  15.98 $\pm$ .08 & 13 	   &  18$\pm$3  & T5.5	    & 14 &    16 & 1.35 \\
0518s28 &      2MASSJ05185995-2828372 &  10 &   5.3166528,  -28.477000 &  15.98 $\pm$ .10 & 13 	   &  19$\pm$3  & T1p	    & 14 &    13 & 1.32 \\
0528s33 &   WISEPJ052844.51-330823.98 &  28 &   5.4790306,  -33.139994 &  16.67 $\pm$ .09 & 28$^a$ &  10$\pm$1  & T8:        & 28 &    12 & 1.12 \\
0542s16 &   WISEPJ054231.27-162829.16 &  28 &   5.7086859,  -16.474767 &  16.58 $\pm$ .14 & 13 	   &  16$\pm$2  & T7-8:      & 28 &    14 & 1.12 \\
0559s14 &      2MASSJ05591914-1404488 &   2 &   5.9886499,  -14.080222 &  13.80 $\pm$ .02 & 13 	   &   7$\pm$1  & T4.5	    & 14 &    13 & 1.44 \\
0611s04 &   WISEPJ061135.13-041024.05 &  28 &   6.1930919,  -4.1733472 &  15.49 $\pm$ .05 & 13 	   &  15$\pm$2  & T0	    & 28 &    15 & 1.21 \\
0612s30 &   WISEPJ061213.89-303612.92 &  28 &   6.2038584,  -30.603589 &  17.10 $\pm$ .19 & 28 	   &  12$\pm$2  & T8:	    & 28 &    13 & 1.21 \\
0623s04 &   WISEPJ062309.93-045624.61 &  28 &   6.3860917,  -4.9401694 &  17.51 $\pm$ .10 & 28 	   &  15$\pm$2  & T8	    & 28 &    12 & 1.21 \\
0627s11 &   WISEPJ062720.07-111428.88 &  28 &   6.4555750,  -11.241356 &  15.49 $\pm$ .05 & 13 	   &  10$\pm$1  & T7	    & 28 &    13 & 1.19 \\
0727n17 &       2MASSIJ0727182+171001 &   5 &   7.4550667,  17.167000 &   15.60 $\pm$ .06 & 13 	   &  10$\pm$1  &  T7	    & 16 &    13 & 1.20 \\
0729s39 &      2MASSJ07290002-3954043 &  17 &   7.4833388,  -39.901222 &  15.92 $\pm$ .08 & 13 	   &   7$\pm$1  & T8pec      & 17 &    15 & 1.21 \\
0751s76 &    WISEPJ075108.79-763449.6 &  28 &   7.8524418,  -76.580444 &  19.34 $\pm$ .05 & 13 	   &  17$\pm$2  & T9	    & 28 &     7 & 0.33 \\
0817s61 &       DENISJ081730.0-615520 &  23 &   8.2916698,  -61.921056 &  13.61 $\pm$ .02 & 13 	   &   5$\pm$1  & T6	    & 23 &     9 & 1.13 \\
0819s03 &   WISEPJ081958.05-033529.01 &  28 &   8.3327913,  -3.5913917 &  14.99 $\pm$ .04 & 13 	   &  13$\pm$2  & T4	    & 28 &    17 & 1.32 \\
0820n10 &     SDSSJ082030.12+103737.0 &  13 &   8.3417025,  10.627000 &   16.98 $\pm$ .19 & 13 	   &  22$\pm$4  & L9.5$\pm$2 & 13 &    14 & 1.32 \\
0830n01 &     SDSSJ083048.80+012831.1 &   9 &   8.5135498,  1.4753056 &   16.29 $\pm$ .11 & 13 	   &  23$\pm$4  & T4.5       & 14 &    13 & 1.32 \\
0926n07 &     ULASJ092624.76+071140.7 &  25 &   9.4402113,  7.1946389 &   17.48 $\pm$ .02 & 30$^a$ &  42$\pm$6  & T3.5       & 25 &    10 & 1.32 \\
0939s24 &      2MASSJ09393548-2448279 &  11 &   9.6598558,  -24.807750 &  15.98 $\pm$ .11 & 13 	   &   7$\pm$1  & T8	    & 14 &    14 & 1.44 \\
0949s15 &      2MASSJ09490860-1545485 &  11 &   9.8190556,  -15.763472 &  16.15 $\pm$ .12 & 13 	   &  22$\pm$3  & T2	    & 14 &    13 & 1.45 \\
0950n01 &              ULASJ0950+0117 &  29 &   9.8464670,  1.2928611 &   18.05 $\pm$ .04 & 30$^a$ &  19$\pm$3  & T8	    & 29 &    15 & 1.46 \\
1007s45 &      2MASSJ10073369-4555147 &  17 &   10.126025,  -45.920753 &  15.65 $\pm$ .07 & 13 	   &  17$\pm$2  & T5	    & 17 &    12 & 1.44 \\
1030n02 &     SDSSJ103026.78+021306.4 &   9 &   10.507439,  2.2182500 &   17.14 $\pm$ .02 & 30$^a$ &  24$\pm$3  & L9.5$\pm$1 & 9	 &    18 & 1.43 \\
1048n09 &     SDSSJ104829.21+091937.8 &  13 &   10.808128,  9.3270278 &   16.59 $\pm$ .15 & 13 	   &  27$\pm$4  & T2.5	    & 13 &    14 & 1.36 \\
1110n01 &    SDSSpJ111010.01+011613.1 &   6 &   11.169447,  1.2702778 &   16.34 $\pm$ .12 & 13 	   &  21$\pm$3  & T5.5	    & 16 &    13 & 1.36 \\
1114s26 &      2MASSJ11145133-2618235 &  11 &   11.247592,  -26.306528 &  15.86 $\pm$ .08 & 13 	   &   9$\pm$1  & T7.5	    & 14 &    12 & 1.38 \\
1122s35 &      2MASSJ11220826-3512363 &  11 &   11.368961,  -35.210083 &  15.02 $\pm$ .04 & 13 	   &  13$\pm$2  & T2	    & 14 &    13 & 1.37 \\
1157n06 &     SDSSJ115700.50+061105.2 &   9 &   11.950136,  6.1847778 &   17.08 $\pm$ .01 & 30$^a$ &  33$\pm$5  &  T1.5      & 14 &    13 & 1.24 \\
1157n09 &     ULASJ115759.04+092200.7 &  22 &   11.966400,  9.3668611 &   16.84 $\pm$ .01 & 30$^a$ &  31$\pm$4  &  T2.5      & 22 &    12 & 1.36 \\
1202n09 &     ULASJ120257.05+090158.8 &  25 &   12.049181,  9.0330000 &   16.80 $\pm$ .01 & 30$^a$ &  28$\pm$4  &  T5        & 25 &    10 & 1.37 \\
1207n02 &     SDSSJ120747.17+024424.8 &   4 &   12.129769,  2.7402500 &   15.58 $\pm$ .07 & 13 	   &  15$\pm$2  &  T0	    & 14 &    13 & 1.37 \\
1209s10 &      2MASSJ12095613-1004008 &   8 &   12.165591,  -10.066889 &  15.91 $\pm$ .08 & 13 	   &  20$\pm$3  & T3	    & 14 &    15 & 1.52 \\
1215s34 &      2MASSJ12154432-3420591 &  17 &   12.262311,  -34.349850 &  16.24 $\pm$ .13 & 13 	   &  23$\pm$4  & T4.5	    & 17 &    16 & 1.52 \\
1300n12 &              ULASJ1300+1221 &  26 &   13.011592,  12.354083 &   16.69 $\pm$ .01 & 30$^a$ &   7$\pm$1  & T8.5       & 25 &    12 & 1.31 \\
1311n01 &    WISEPJ131106.24+012252.4 &  28 &   13.185066,  1.3812222 &   19.16 $\pm$ .12 & 28 	   &  16$\pm$2  & T9:	    & 28 &    10 & 0.47 \\
1402n08 &     SDSSJ140255.66+080055.2 &  13 &   14.048789,  8.0153611 &   16.84 $\pm$ .18 & 13 	   &  29$\pm$5  & T1.5	    & 13 &    17 & 1.45 \\
1404s31 &      2MASSJ14044941-3159329 &  17 &   14.080392,  -31.992517 &  15.58 $\pm$ .06 & 13 	   &  17$\pm$2  & T2.5	    & 17 &    13 & 1.45 \\
1459n08 &     ULASJ145935.25+085751.2 &  25 &   14.993125,  8.9642222 &   17.94 $\pm$ .03 & 30$^a$ &  50$\pm$7  & T4.5       & 25 &    11 & 1.44 \\
1504n10 &     SDSSJ150411.63+102718.4 &  13 &   15.069878,  10.455422 &   16.50 $\pm$ .01 & 30$^a$ &  15$\pm$2  & T7	    & 13 &    11 & 1.31 \\
1511n06 &     SDSSJ151114.66+060742.9 &  13 &   15.187406,  6.1286389 &   16.02 $\pm$ .08 & 13 	   &  19$\pm$3  & T0$\pm$2   & 13 &    10 & 1.44 \\
1521n01 &     SDSSJ152103.24+013142.7 &   9 &   15.350908,  1.5285000 &   16.40 $\pm$ .10 & 13 	   &  25$\pm$4  & T2:	    & 14 &    10 & 1.14 \\
\end{tabular}                                                                             	      
\end{minipage}                                                                            	      
\end{table*}                                                                              	      
\begin{table*}                                                                            	      
\begin{minipage}{216mm}                                                                   	      
\contcaption{}                                                                            	      
\begin{tabular}{lllllllllll}                                        	      			      
Short   & ~~~~~~ Discovery         & Ref &~~~ Discovery $\alpha,\delta$ & ~~J $\pm$ $\sigma$ & Ref & d$_{p} \pm  \sigma$& ~IR & Ref &N$_E$ & $\Delta$T  \\
Name    & ~~~~~~~~ Name              & Dis.&~~~~~  hrs,deg J2000      &                     & ~J     &   ~~pc   & SpT       & SpT &      &   yr    \\ 
 1553n15  &        2MASSIJ1553022+153236 &  5 &   15.883966, 15.543583 &    15.82 $\pm$ 0.07    & 13  	 &  11$\pm$2 & T7       & 14 &    11 & 1.29 \\ 
 1615n13  &       2MASSJ16150413+1340079 & 17 &   16.251146, 13.668869 &    16.35 $\pm$ 0.09    & 13  	 &  19$\pm$3 & T6       & 17 &    11 & 1.38 \\ 
 1617n18  &     WISEPJ161705.75+180714.0 & 28 &   16.284929, 18.120556 &    17.66 $\pm$ 0.08    & 28$^a$ &  16$\pm$2 & T8       & 28 &     9 & 1.18 \\ 
 1630n08  &      SDSSJ163022.92+081822.0 & 13 &   16.506374, 8.3061389 &    16.40 $\pm$ 0.11    & 13  	 &  22$\pm$3 & T5.5     & 13 &    11 & 1.38 \\ 
 1741n25  &              WISEPJ1741+2553 & 28 &   17.690071, 25.888753 &    16.45 $\pm$ 0.10    & 28  	 &   4$\pm$1 &  T9      & 28 &    10 & 1.20 \\ 
 1750n17  &     SDSSpJ175032.96+175903.9 &  6 &   17.842480, 17.984500 &    16.34 $\pm$ 0.10    & 13  	 &  25$\pm$4 &  T3.5    & 14 &    13 & 1.37 \\ 
 1812n27  &     WISEPJ181210.85+272144.3 & 28 &   18.203014, 27.362306 &    18.19 $\pm$ 0.06    & 28$^a$ &  15$\pm$2 &  T8.5    & 28 &     7 & 1.18 \\ 
 1821n14  &       2MASSJ18212815+1414010 & 24 &   18.357819, 14.233611 &    13.43 $\pm$ 0.02    & 24  	 &  10$\pm$1 &  L4.5    & 24 &     7 & 1.20 \\ 
 1828s48  &       2MASSJ18283572-4849046 &  8 &  18.476589, -48.817944 &   15.18 $\pm$ 0.06    & 13      &  12$\pm$2 & T5.5     & 14 &    14 & 1.90 \\ 
 1934s21  &          CFBDSJ193430-214221 & 19 &  19.575111, -21.705833 &   16.77 $\pm$ 0.15    & 13      &  30$\pm$5 & T3.5     & 19 &    13 & 1.90 \\ 
 1936s55  &       2MASSJ19360187-5502322 & 20 &  19.600519, -55.042278 &   14.49 $\pm$ 0.04    & 13      &  16$\pm$2 & L5       & 20 &    13 & 1.18 \\ 
 2018s74  &     WISEPJ201824.98-742326.1 & 28 &  20.306938, -74.390581 &   17.10 $\pm$ 0.30     & b      &  23$\pm$5 & T7       & 28 &    12 & 1.36 \\ 
 2043s15  &      SDSSJ204317.69-155103.4 & 13 &  20.721581, -15.850861 &   16.62 $\pm$ 0.16    & 13      &  21$\pm$3 & L9       & 13 &    12 & 1.90 \\ 
 2047s07  &      SDSSJ204749.61-071818.3 &  9 &  20.797108, -7.3048889 &   16.95 $\pm$ 0.20    & 13      &  29$\pm$5 & T0:      & 14 &    12 & 1.90 \\ 
 2052s16  &      SDSSJ205235.31-160929.8 & 13 &  20.876430, -16.158556 &   16.33 $\pm$ 0.12    & 13      &  23$\pm$3 & T1$\pm$1 & 13 &    12 & 1.90 \\ 
 2124n01  &      SDSSJ212413.89+010000.3 &  9 &  21.403852, 0.99997222 &   16.03 $\pm$ 0.07    & 13      &  20$\pm$3 & T5       & 14 &    13 & 1.90 \\ 
 2139n02  &       2MASSJ21392676+0220226 & 20 &   21.657436, 2.3396389 &    15.26 $\pm$ 0.05    & 13     &  14$\pm$2 &  T1.5    & 14 &     9 & 1.90 \\ 
 2151s48  &       2MASSJ21513839-4853542 & 12 &  21.860664, -48.898389 &   15.73 $\pm$ 0.08    & 13      &  19$\pm$3 & T4       & 14 &     9 & 1.90 \\ 
 2154s10  &       2MASSJ21542494-1023022 & 17 &  21.906927, -10.383950 &   16.42 $\pm$ 0.12    & 13      &  25$\pm$4 & T4.5     & 17 &    11 & 1.90 \\ 
 2228s43  &       2MASSJ22282889-4310262 &  7 &  22.474691, -43.173942 &   15.66 $\pm$ 0.07    & 13      &  14$\pm$2 & T6       & 14 &    11 & 1.90 \\ 
 2229n01  &      ULASJ222958.30+010217.2 & 21 &   22.499527, 1.0381111 &    17.88 $\pm$ 0.04    & 30$^a$ &  50$\pm$7 &  T2.5    & 21 &    11 & 1.90 \\ 
 2239n16  &    WISEPJ223937.55+161716.20 & 28 &   22.660431, 16.287833 &    16.08 $\pm$ 0.08    & 13  	 &  22$\pm$3 &  T3      & 28 &     8 & 1.20 \\ 
 2325s41  &    WISEPJ232519.54-410534.90 & 28 &  23.422094, -41.093028 &   19.75 $\pm$ 0.05    & 28$^a$  &  20$\pm$3 & T9       & 28 &     7 & 1.18 \\ 
 2331s47  &       2MASSJ23312378-4718274 &  8 &  23.523272, -47.307608 &   15.66 $\pm$ 0.07    & 13 	 &  17$\pm$2 & T5       & 14 &     9 & 1.90 \\ 
 2342n08  &               ULASJ2342+0856 & 29 &   23.708048, 8.9389167 &    16.37 $\pm$ 0.01    & 30$^a$ &  17$\pm$2 &  T6.5    & 29 &    13 & 1.90 \\ 
 2356s15  &        2MASSIJ2356547-155310 &  5 &  23.948547, -15.886417 &   15.82 $\pm$ 0.06    & 13 	 &  17$\pm$2 & T5.5     & 14 &     8 & 1.67 \\ 
\end{tabular}               
\end{minipage}              
J= J band magnitude, $\Delta$T = epoch range, d$_{p}$ = photometric distance, N$_E$ = Number of epochs todate.\newline
Ref: 1-\cite{1999Sci...283.1718M},
 2-\cite{2000AJ....120.1100B},
 3-\cite{2000AJ....120..447K},
 4-\cite{2002AJ....123.3409H},
 5-\cite{2002ApJ...564..421B},
 6-\cite{2002ApJ...564..466G},
 7-\cite{2003AJ....126.2487B},
 8-\cite{2004AJ....127.2856B},
 9-\cite{2004AJ....127.3553K},
10-\cite{2004ApJ...604L..61C},
11-\cite{2005AJ....130.2326T},
12-\cite{2005AJ....130.2347E},
13-\cite{2006AJ....131.2722C},
14-\cite{2006ApJ...637.1067B},
15-\cite{2006ApJ...651L..57A},
16-\cite{2006ApJS..166..585B},
17-\cite{2007AJ....134.1162L},
18-\cite{2007MNRAS.379.1423L},
19-\cite{2008AA...482..961D}',
20-\cite{2008AJ....136.1290R},
21-\cite{2008MNRAS.385L..53C},
22-\cite{2008MNRAS.390..304P},
23-\cite{2010ApJ...718L..38A},
24-\cite{2010ApJS..190..100K},
25-\cite{2010MNRAS.404.1952B},
26-\cite{2010MNRAS.405.1140G},
27-\cite{2011AA...532L...5S},
28-\cite{2011ApJS..197...19K},
29-\cite{burningham2013},
30-\cite{2006AJ....131.1163S} \newline
Ref J Notes: a=magnitude on MKO system, b=estimated from first NPARSEC observation.
\end{table*}